\documentclass[12pt]{iopart}

\usepackage{epsfig}
\usepackage{iopams}
\begin{document}
\title{Compressible Sherrington-Kirkpatrick spin-glass model}

\author{Danilo B Liarte, Silvio R Salinas and Carlos S O Yokoi}

\address{Instituto de F\'{\i}sica, Universidade de S\~{a}o Paulo,
  Caixa Postal 66318, 05315-970 S\~{a}o Paulo, SP, Brazil}

\ead{danilo@if.usp.br}

\begin{abstract}
We introduce a Sherrington-Kirkpatrick spin-glass model with the addition of
elastic degrees of freedom. The problem is formulated in terms of an
effective four-spin Hamiltonian in the pressure ensemble, which can be
treated by the replica method. In the replica-symmetric approximation, we
analyze the pressure-temperature phase diagram, and obtain expressions for
the critical boundaries between the disordered and the ordered (spin-glass
and ferromagnetic) phases. The second-order para-ferromagnetic border ends
at a tricritical point, beyond which the transition becomes discontinuous.
We use these results to make contact with the temperature-concentration
phase diagrams of mixtures of hydrogen-bonded crystals.
\end{abstract}



\maketitle

\section{Introduction}
Orientational models with disorder and elastic degrees of freedom have been
used to investigate phase transitions in a number of systems, as the mixed
molecular crystal K(CN)$_x$Br$_{1-x}$ \cite{tadic94}. The work on
quadrupolar glass models in the presence of random strain fields, which are
supposed to mimic the random mismatching of molecular groups, was reviewed
by Binder and Reger \cite{binder92}. A compressible spherical model, with
random bonds and in random fields, has been introduced to account for the
peculiar behavior of compositionally disordered perovskites, also known as
relaxor ferroelectrics \cite{blinc01}. Disordered Ising models, with
random competing interactions and elastic degrees of freedom, have also been
used to account for the phase diagram and the glassy transition in mixtures
of ferro and antiferroelectric hydrogen-bonded crystals of the KDP family 
\cite{courtens83,matsushita85,papantopoulos94}. More recently, random quadrupolar models
in the presence of anisotropic strain fields have been studied in the
context of nematic liquid crystal elastomers \cite{petridis06}. These
investigations on disordered compressible models provided the motivation to
introduce a simple compressible Sherrington-Kirkpatrick (SK) Ising
spin-glass model \cite{sherrington75,binder86}, and use standard techniques to obtain
the global phase diagram in terms of pressure and temperature.

The investigation of compressible Ising models has a long history \cite%
{domb56,salinas75,bergman76}. Mean-field and renormalization-group calculations indicate that
the inclusion of elastic degrees of freedom may change the nature of the
continuous Ising transition. An Ising antiferromagnet on the triangular
lattice, with properly chosen elastic degrees of freedom, is known to
provide a mechanism to explain the formation of striped phases \cite%
{chen86,gu96}. Although elastic spin models with uniform interactions have been
much investigated, disordered compressible spin models are less explored.
Nowadays it has been feasible to carry out detailed computer simulations for
compressible models \cite{tavazza04,zhu05,landau06}, including disordered compressible models,
as in the work of Marshall for a compressible Edwards-Anderson Ising
spin-glass on a two-dimensional lattice \cite{marshall06,marshall07}. These works have
provided further motivation to study the phase diagram of the simple
compressible SK spin-glass model.

We treat the mean-field SK system in the pressure ensemble, which is more
adequate to analyze the phase diagram in terms of the intensive field
variables, temperature $T$ and pressure $p$. Using the replica method, the
calculation of the free energy is reduced to the minimization of a
functional of three sets of replica variables \cite{sherrington75}. We
obtain a number of results in the replica-symmetric approximation. In
particular, we show that the presence of magneto-elastic couplings
introduces a simple shift in the border between paramagnetic and spin-glass
phases, and that there appears a tricritical point along the
para-ferromagnetic border. Also, we perform an analysis of stability of the
replica-symmetric solution and locate the de Almeida-Thouless line \cite{almeida78}. In the
Conclusions, we consider a compressible two-sublattice SK model to make
contact with the experimental phase diagrams of mixtures of ferro and
antiferroelectric hydrogen-bonded crystals.

\section{The compressible SK model}

The compressible Sherrington-Kirkpatrick spin-glass model is defined by the
Hamiltonian%
\begin{equation}
\mathcal{H}=-\sum_{1\leq i<j\leq N}J_{ij}\left[ 1-\gamma \left( v-v_{0}\right) \right]
S_{i}S_{j}-H\sum_{i=1}^{N}S_{i}+\frac{1}{2}kN\left( v-v_{0}\right) ^{2},
\end{equation}%
where $S_{i}=\pm 1$, with $i=1,...,N$, $v$ is the specific volume, $v_{0}>0$
is a constant parameter, $\gamma $ is the magneto-elastic coupling, $H$ is
the external field, and $k>0$ is a uniform elastic constant. As usual, $%
\left\{ {\large J_{ij}}\right\} $ is a set of independent and identically
distributed random variables, with suitably scaled mean values, $%
\left\langle J_{ij}\right\rangle =J_{0}/N$, and variances, $\left\langle
\left( J_{ij}-\left\langle J_{ij}\right\rangle \right) ^{2}\right\rangle
=J^{2}/N$. Given a configuration $\left\{ J_{ij}\right\} $, we write the
partition function in the pressure ensemble,%
\begin{equation}
Y=\int_{0}^{\infty }dv\exp \left( -\beta pvN\right) \mathbf{Tr}\exp \left(
-\beta \mathcal{H}\right) ,
\end{equation}%
where $p$ is the pressure, $\beta =1/\left( k_{B}T\right) $, and the trace
is a sum over spin configurations. Performing the volume integration, and
discarding irrelevant terms in the thermodynamic limit, we have 
\begin{equation}
Y=\exp \left[ -\beta N\left( pv_{0}-\frac{p^{2}}{2k}\right) \right] \mathbf{%
Tr}\exp \left( -\beta \mathcal{H}_{eff}\right) ,
\end{equation}%
with an effective four-spin Hamiltonian,%
\begin{equation}
\mathcal{H}_{eff}=-\left( 1+\frac{\gamma p}{k}\right)
\sum_{i<j}J_{ij}S_{i}S_{j}-\frac{\gamma ^{2}}{2kN}\left(
\sum_{i<j}J_{ij}S_{i}S_{j}\right) ^{2}-H\sum_{i}S_{i}.
\label{heff}
\end{equation}%
Using a Gaussian identity, we write the partition function as%
\begin{eqnarray}
\fl Y=\exp \left[ -\beta N\left( pv_{0}-\frac{p^{2}}{2k}\right) \right]
\int_{-\infty }^{+\infty }dx\exp \left( -\frac{N\beta \gamma ^{2}J^{2}}{2k}%
x^{2}\right) 
\nonumber \\
\times \mathbf{Tr}\exp \left[ \beta \left( 1+\frac{\gamma p}{k}+\frac{%
J\gamma ^{2}x}{k}\right) \sum_{i<j}J_{ij}S_{i}S_{j}+\beta H\sum_{i}S_{i}%
\right] ,
\end{eqnarray}
which is in a more convenient form to be treated by replicas.

According to the replica method,%
\begin{equation}
\left\langle \ln Y\right\rangle =\lim_{n\rightarrow 0}\frac{1}{n}\ln
\left\langle Y^{n}\right\rangle ,
\end{equation}%
where%
\begin{eqnarray}
\fl \left\langle Y^{n}\right\rangle =\exp \left[ -\beta Nn\left( pv_{0}-\frac{%
p^{2}}{2k}\right) \right] \int_{-\infty }^{+\infty }\prod_{\alpha
=1}^{n}dx^{\alpha }\exp \left[ -\frac{N\beta \gamma ^{2}J^{2}}{2k}%
\,\sum_{\alpha }\left( x^{\alpha }\right) ^{2}\right] 
\nonumber \\
\times \left\langle \mathbf{Tr}_{n}\exp \left[ \sum_{i<j}J_{ij}\left( \beta
\sum_{\alpha }\xi ^{\alpha }S_{i}^{\alpha }S_{j}^{\alpha }\right) +\beta
H\sum_{i,\alpha }S_{i}^{\alpha }\right] \right\rangle ,
\end{eqnarray}%
with the definition%
\begin{equation}
\xi ^{\alpha }=1+\frac{\gamma p}{k}+\frac{J\gamma ^{2}x^{\alpha }}{k}.
\end{equation}%
Taking the average over the exchange configurations, and discarding
irrelevant terms in the thermodynamic limit, we have%
\begin{eqnarray}
\fl \left\langle Y^{n}\right\rangle =\exp \left[ -\beta Nn\left( pv_{0}-\frac{%
p^{2}}{2k}\right) \right] \int_{-\infty }^{+\infty }\prod_{\alpha
}dx^{\alpha }\exp \left[ -\frac{N\beta \gamma ^{2}}{2k}\,\sum_{\alpha
}\left( x^{\alpha }\right) ^{2}\right] 
\nonumber \\
\fl \times \mathbf{Tr}_{n}\exp \left[ \frac{\beta ^{2}J^{2}}{2N}\sum_{\alpha
,\beta }\xi ^{\alpha }\xi ^{\beta }\sum_{i<j}S_{i}^{\alpha }S_{i}^{\beta
}S_{j}^{\alpha }S_{j}^{\beta }+\frac{\beta J_{0}}{N}\sum_{\alpha }\xi
^{\alpha }\sum_{i<j}S_{i}^{\alpha }S_{j}^{\alpha }
+\beta H\sum_{i,\alpha
}S_{i}^{\alpha }\right] ,
\end{eqnarray}%
which can be rewritten as%
\begin{eqnarray}
\fl \left\langle Y^{n}\right\rangle =\exp \left[ -\beta Nn\left( pv_{0}-\frac{%
p^{2}}{2k}\right) \right] \int_{-\infty }^{+\infty }\prod_{\alpha }dx^{\alpha }\exp \left\{
\,\sum_{\alpha }\left[ -\frac{N\beta J^{2}\gamma ^{2}}{2k}\left( x^{\alpha
}\right) ^{2}
\nonumber \right. \right. \\ \left. \left.
+\frac{N\beta ^{2}J^{2}}{4}\left( \xi ^{\alpha }\right) ^{2}%
\right] \right\} \mathbf{Tr}_{n}\exp \left\{ \frac{\beta ^{2}J^{2}}{2N}\sum_{\alpha
<\beta }\xi ^{\alpha }\xi ^{\beta }\left( \sum_{i}S_{i}^{\alpha
}S_{i}^{\beta }\right) ^{2}
\nonumber \right. \\ \left.
+\frac{\beta J_{0}}{2N}\sum_{\alpha }\xi ^{\alpha
}\left( \sum_{i}S_{i}^{\alpha }\right) ^{2}+\beta H\sum_{i,\alpha
}S_{i}^{\alpha }\right\} .
\end{eqnarray}

We now use Gaussian identities to introduce the new set of variables $%
\left\{ m_{\alpha }\right\} $, associated with the magnetization, and $%
\left\{ q_{\alpha \beta }\right\} $, associated with the overlap between
replicas, so that 
\begin{equation}
\fl \left\langle Y^{n}\right\rangle =\int_{-\infty }^{+\infty }\prod_{\alpha
}dx^{\alpha }\int_{-\infty }^{+\infty }\prod_{\alpha }dm^{\alpha
}\int_{-\infty }^{+\infty }\prod_{\alpha <\beta }dq^{\alpha \beta }\exp 
\left[ NG\left( x^{\alpha },m^{\alpha },q^{\alpha \beta }\right) \right] ,
\end{equation}%
where%
\begin{eqnarray}
\fl G\left( x^{\alpha },m^{\alpha },q^{\alpha \beta }\right) =-\beta n\left(
pv_{0}-\frac{p^{2}}{2k}\right) -\frac{\beta \gamma ^{2}J^{2}}{2k}%
\,\sum_{\alpha }\left( x^{\alpha }\right) ^{2}+\frac{\beta ^{2}J^{2}}{4}%
\sum_{\alpha }\left( \xi ^{\alpha }\right) ^{2}
\nonumber \\
-\frac{\beta ^{2}J^{2}}{2}\sum_{\alpha <\beta }\xi ^{\alpha }\xi ^{\beta
}\left( q^{\alpha \beta }\right) ^{2}-\frac{\beta J_{0}}{2}\sum_{\alpha }\xi
^{\alpha }\left( m^{\alpha }\right) ^{2}
\nonumber \\
+\ln \mathrm{Tr}\exp \left( \beta ^{2}J^{2}\sum_{\alpha <\beta }\xi ^{\alpha
}\xi ^{\beta }q^{\alpha \beta }S^{\alpha }S^{\beta }+\beta J_{0}\sum_{\alpha
}\xi ^{\alpha }m^{\alpha }S^{\alpha }+\beta H\sum_{\alpha }S^{\alpha
}\right) .
\end{eqnarray}%
In the thermodynamic limit, the free energy comes from 
\begin{equation}
g=g(T,p,H)=-\frac{1}{\beta }\lim_{n\rightarrow 0}\frac{1}{n}\max G\left(
x^{\alpha },m^{\alpha },q^{\alpha \beta }\right) ,
\end{equation}%
where the maximum should be taken with respect to the three sets of replica
variables.

\section{Replica-symmetric solution}

In the replica-symmetric solution, with $x_{\alpha }=x$ and $m_{\alpha }=m$,
for all $\alpha $, and $q_{\alpha \beta }=q$, for all pairs $\left( \alpha
\beta \right) $, we have 
\begin{eqnarray}
\fl -\beta g=-\beta \left( pv_{0}-\frac{p^{2}}{2k}\right) -\frac{\beta \gamma
^{2}J^{2}}{2k}\,x^{2}+\frac{\beta ^{2}J^{2}}{4}\,\xi ^{2}\left( 1-q\right)
^{2}-\frac{\beta J_{0}\xi }{2}m^{2} 
\nonumber \\
+\int_{-\infty }^{+\infty }\frac{dz}{\sqrt{2\pi }}\exp \left( -\frac{z^{2}}{2%
}\right) \ln 2\cosh \left[ \beta \left( J\xi q^{1/2}z+J_{0}\xi m+H\right) %
\right] ,
\end{eqnarray}%
with%
\begin{equation}
\xi =1+\frac{\gamma p}{k}+\frac{J\gamma ^{2}x}{k}.  \label{def}
\end{equation}%
The equations of state are given by%
\begin{equation}
q=\int_{-\infty }^{+\infty }\frac{dz}{\sqrt{2\pi }}\exp \left( -\frac{z^{2}}{%
2}\right) \tanh ^{2}\left[ \beta \left( J\xi q^{1/2}z+J_{0}\xi m+H\right) %
\right] ,  \label{q}
\end{equation}

\begin{equation}
m=\int_{-\infty }^{+\infty }\frac{dz}{\sqrt{2\pi }}\exp \left( -\frac{z^{2}}{%
2}\right) \tanh \left[ \beta \left( J\xi q^{1/2}z+J_{0}\xi m+H\right) \right]
,  \label{m}
\end{equation}%
and%
\begin{equation}
x=\frac{\beta J}{2}\,\xi \left( 1-q^{2}\right) +\frac{J_{0}}{2J}\,m^{2}.
\label{x}
\end{equation}

We now restrict the analysis to zero field ($H=0$).

The transition between the spin-glass and the paramagnetic phase comes from
the expansions of equations (\ref{q}) and (\ref{x}), 
\[
q=\frac{1}{2\beta ^{2}J^{2}\xi ^{2}}\left( 1-\frac{1}{\beta ^{2}J^{2}\xi ^{2}%
}\right) +O(q^{2}), 
\]%
and 
\begin{equation}
x=\frac{\beta J\xi }{2}+O(q^{2}).
\end{equation}%
There is a (positive) solution, $q>0$, for%
\begin{equation}
1-\frac{1}{\xi \beta J}>0.
\end{equation}%
Introducing the notation%
\begin{equation}
t=\frac{1}{\beta J}=\frac{k_{B}T}{J},
\end{equation}%
and using the definition of $\xi $, given by Eq. (\ref{def}), we have the
second-order boundary between the paramagnetic and the spin-glass phases,%
\begin{equation}
t_{c1}=1+\frac{\gamma p}{k}+\frac{J\gamma ^{2}}{2k}.
\end{equation}

In zero field, the calculation of the para-ferromagnetic transition comes
from the analysis of the expansion%
\begin{equation}
x=\frac{\beta J}{2}\,\xi +\frac{J_{0}}{2J}m^{2}+O(q^{2}),
\end{equation}%
which can be written as%
\begin{equation}
\xi \left( 1-\frac{\beta J^{2}\gamma ^{2}}{2k}\right) =\left( 1+\frac{\gamma
p}{k}\right) +\frac{J_{0}\gamma ^{2}}{2k}m^{2}+O(q^{2}).  \label{xx}
\end{equation}%
Introducing the dimensionless and more compact notation%
\begin{equation}
j_{0}=\frac{J_{0}}{J},\qquad a=1+\frac{\gamma p}{k},\qquad b=\frac{J\gamma
^{2}}{2k},
\end{equation}%
Eq. (\ref{xx}) can be rewritten as%
\[
\xi \left( 1-\frac{b}{2t}\right) =a+bj_{0}m^{2}+O(q^{2}). 
\]%
Inserting this expansion into the equations of state, it is easy to show that%
\begin{equation}
1=\frac{aj_{0}}{t-b}\left( 1+\frac{j_{0}b}{a}\,m^{2}\right) -\frac{1}{3}%
\left( \frac{a}{t-b}\right) ^{3}\left( j_{0}^{3}m^{2}+3qj_{0}\right)
+O(m^{3},q^{3/2}),
\end{equation}%
from which we have the second-order para-ferromagnetic border,%
\begin{equation}
t_{c2}=aj_{0}+b=\left( 1+\frac{\gamma p}{k}\right) j_{0}+\frac{J\gamma ^{2}}{%
2k}.
\end{equation}

The transition between the spin-glass and the ferromagnetic phases may be
calculated from an analysis of the zero-field susceptibility. It is not
difficult to show that this border is given by 
\begin{equation}
t_{c3}=b\left( 1-q^{2}\right) +aj_{0}\left( 1-q\right) ,
\end{equation}%
where $q$ comes from the equation%
\begin{equation}
q=\int_{-\infty }^{+\infty }\frac{dz}{\sqrt{2\pi }}\exp \left( -\frac{z^{2}}{%
2}\right) \tanh ^{2}\left[ \frac{aq^{1/2}z}{t-b\left( 1-q^{2}\right) }\right]
.
\end{equation}

In order to investigate the eventual existence of tricritical points, we
write the expansions%
\begin{eqnarray}
\fl m=\frac{j_{0}am}{t-b}\left( 1+\frac{b}{a}j_{0}m^{2}\right) 
-\frac{1}{3}\frac{j_{0}^{3}a^{3}m}{\left( t-b\right) ^{3}}\left(
m^{2}+3qj_{0}^{-2}+3\frac{b}{a}j_{0}m^{4}+9\frac{b}{a}j_{0}^{-1}qm^{2}%
\right)
\nonumber \\
 +O\left( m^{6},q^{3}\right) ,  \label{m2}
\end{eqnarray}%
and%
\begin{eqnarray}
\fl q=\left( \frac{j_{0}a}{t-b}\right) ^{2}\left( 1+2\frac{b}{a}%
j_{0}m^{2}\right) \left( qj_{0}^{-2}+m^{2}\right) -\frac{2}{3}\left( \frac{j_{0}a}{t-b}\right) ^{4}\left(
3q^{2}j_{0}^{-4}+m^{4}
\nonumber \right. \\ \left.
+6qj_{0}^{-2}m^{2}\right) +O\left( m^{6},q^{3}\right) .
\label{q2}
\end{eqnarray}%
For $m=0$, the coefficients of Eq. (\ref{q2}) indicate that there is no
possibility of a tricritical point along the second-order border between the
paramagnetic and the spin-glass phases. At the para-ferromagnetic border,
however, it is not difficult to locate a tricritical point at%
\begin{equation}
p_{tc}=\frac{3J\gamma j_{0}\left( j_{0}^{2}-1\right) }{2\left(
j_{0}^{2}+2\right) }-\frac{k}{\gamma }.
\end{equation}

Let us turn to the Hessian matrix associated with the replica-symmetric
solutions of $-\beta g$. There is a total of thirteen distinct elements in the $n \rightarrow 0$ limit, seven of which coming from derivatives involving the $x^{\alpha}$ variables. As in the standard SK
model, the paramagnetic phase is always stable, and the eigenvector subspace is generated by three classes of vectors, which have the general form
\begin{eqnarray}
\vec{\mu} = \left( \begin{array}{c}
\left\{\phi^\alpha\right\} \\
\left\{\epsilon^\alpha\right\} \\
\left\{\eta^{(\alpha\beta)}\right\}
\end{array} \right), \quad \alpha, \beta = 1, \cdots, n,
\end{eqnarray}
where $\{\phi^\alpha\}$, $\{\epsilon^\alpha\}$ and $\{\eta^{(\alpha \beta)}\}$ are column-vectors with $n$, $n$ and $n(n-1)/2$ elements respectively \cite{almeida78}.
The eigenvalue spectrum is obtained straightforwardly following the same steps of the rigid model. In the ordered region, the
change of sign of the transverse eigenvalue leads to an instability. The de
Almeida-Thouless line is given by%
\begin{eqnarray}
\fl \left( \frac{k_{B}T}{J}\right) ^{2}=\left( a+2bx\right) ^{2}\int_{-\infty
}^{+\infty }\frac{dz}{\sqrt{2\pi }}\exp \left( -\frac{z^{2}}{2}\right) \mathrm{%
sech}^{4}\left[ \beta \left( a+2bx\right) \left( Jq^{1/2}z+J_{0}m\right) %
\right] ,
\end{eqnarray}%
where $H=0$, and $m$, $q$, and $x$, are the equations of state.

\section{Conclusions}

\begin{figure}[!h]
  \vspace{1.5cm}
  \begin{center}
    \epsfig{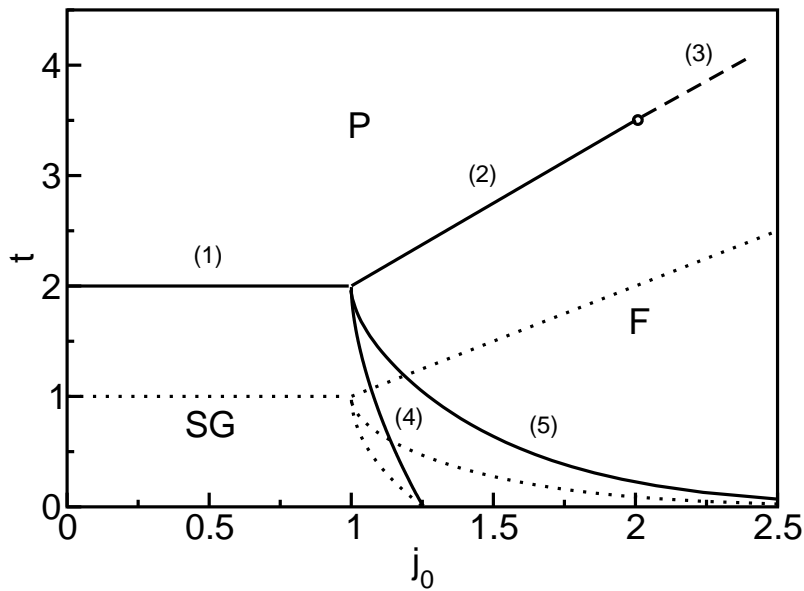}
    \vspace{0.1 cm}
    \caption{Comparison between the $t-j_{0}$
phase diagrams of the rigid (dotted lines) and compressible (solid lines) SK
models. We indicate the paramagnetic (P), spin-glass (SG), and ferromagnetic
(F) phases. Solid lines (1) to (4) are the boundaries coming from the
replica-symmetric solution of the compressible SK model. Solid line (5) is
the AT instability border. There is a tricritical point separating the
continuous (solid line) and first-order (dashed line) P-F border of the
compressible SK model.}
    \label{figure1}
  \end{center}
\end{figure}

It is interesting to draw a typical phase diagram in terms of temperature 
$t=k_{B}T/J$ and the ratio $j_{0}=J_{0}/J$ (in zero field). The dotted lines
in Figure \ref{figure1} are the well-known\ replica-symmetric second-order boundaries,
with the addition of the de Almeida-Thouless (AT) instability line,
for the rigid SK model (with no magneto-elastic coupling). The location of
these borders and of the AT line is shifted in the presence of a
magneto-elastic coupling. The spin-glass and ferromagnetic regions are broadened as a result of a decrease of the effective energy of the system due to the magneto-elastic coupling, as seen in equation (\ref{heff}). The solid and dashed lines were drawn for the
parameter values $a=1+\gamma p/k=1.5$ and $b=J\gamma ^{2}/2k=0.5$. For these
particular values of pressure and elastic parameters, the second-order
boundary between the paramagnetic and the ferromagnetic phases, given by 
$t=1/2+3j_{0}/2$, ends at a tricritical point, at $j_{0}=2$, beyond which
there is a (dashed) line of first-order transitions. These are typical
results for the compressible SK model in the replica-symmetric approximation.

With the exception of the well-known corrections in the spin-glass region 
\cite{binder86}, in analogy to the treatment of the rigid SK model, the
form of the effective four-spin Hamiltonian in the pressure ensemble does
not indicate any drastic qualitative changes of the phase diagrams even if
we go beyond the simple replica-symmetric approximation.

\begin{figure}[!h]
  \vspace{1.5cm}
  \begin{center}
    \epsfig{file=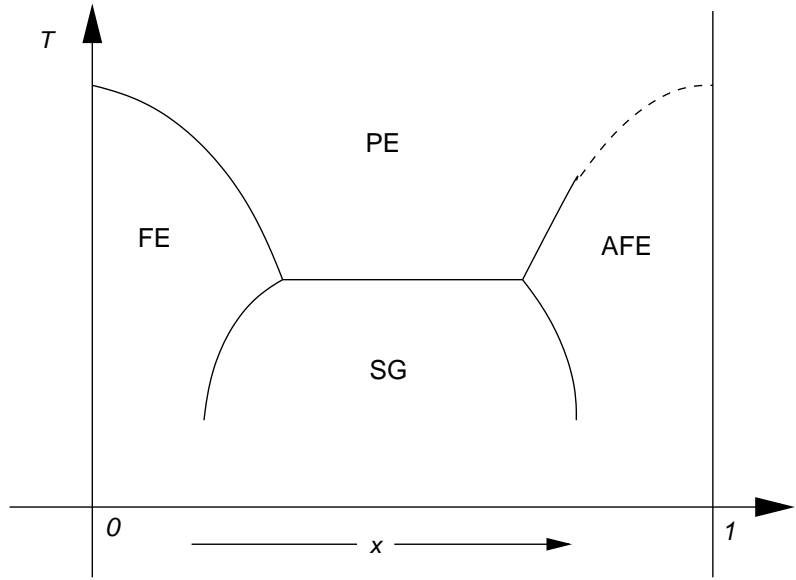,scale=0.55}
    \vspace{0.1 cm}
    \caption{Schematic phase diagram for the
two-sublattice compressiblke SK model for RADP in terms of temperature $T$
and composition $x$. We indicate paraelectric, ferroelectric,
antiferroelectric, and glassy regions. Solid lines represent continuous
transitions. The dashed line is a first-order transition.}
    \label{figure2}
  \end{center}
\end{figure}

We now make contact with previous work for mixtures of ferroelectric and
antiferroelectric hydrogen-bonded crystals of the KDP family. A disordered
Ising model in a random field, with the inclusion of elastic degrees of
freedom, has been used to investigate the glassy transition in 
Rb$_{1-x}$(NH$_4$)$_{x}$H$_{2}$AsO$_{4}$, known as RADA \cite{papantopoulos94}. Also, the phase diagram of Rb$_{1-x}$(NH$_{4}$)$_{x}$H$_{2}$PO$_{4}$ (RADP), in terms of temperature $T$ and concentration $x$,
has been drawn on the basis of a cluster calculation for a model of Ising
pseudo-spins \cite{courtens83,matsushita85}. The Rubidium crystals, RbH$_{2}$PO$_{4}$
(RDP) and RbH$_{2}$AsO$_{4}$ (RDA), display a ferroelectric transition, at
about $150$ $K$, similar to the well-known ferroelectric phase transition in
the isomorphous KH$_{2}$PO$_{4}$ (KDP) crystals. The Ammonium crystals, NH$_{4}$H$_{2}$AsO$_{4}$ (ADA) and NH$_{4}$H$_{2}$PO$_{4}$ (ADP), display a strong
first-order antiferroelectric transition. At the mean-field level, the
description of the $T-x$ phase diagrams requires the consideration of two
sublattices. We thus use a two-sublattice compressible SK model, given by
the Hamiltonian
\begin{eqnarray}
\mathcal{H}=-\sum_{i\in A,j\in B} J_{ij}\left[ 1-\gamma \left( v-v_{0}\right) \right] S_{i}S_{j}+\frac{1}{2}kN\left( v-v_{0}\right) ^{2},
\end{eqnarray}
where the sum refers to all distinct pairs of spins belonging to different
sublattices, $A$ and $B$. The random variables $\left\{ J_{ij}\right\} $ are
associated with a Gaussian distribution, with suitably scaled moments, $\left\langle J_{ij}\right\rangle =J_{0}/N$, and $\left\langle \left(
J_{ij}-\left\langle J_{ij}\right\rangle \right) ^{2}\right\rangle =J^{2}/N$,
where we choose
\begin{equation}
J_{0}=xJ_{R}-\left( 1-x\right) J_{A},
\end{equation}
and make $J^{2}\rightarrow J^{2}x\left( 1-x\right) $, in order to mimic
ferroelectric ($x=1$) and antiferroelectric ($x=0$) transitions.
Two-sublattice Sherrington-Kirkpatrick models, even in the presence of
random fields, have been studied in connection with transitions from
antiferromagnetic to spin-glass phases \cite{korenblit85,fyodorov87a,fyodorov87b,vieira00}. The
addition of elastic degrees of freedom, according to the steps of the
previous sections, suggests a typical phase diagram as sketched in Figure \ref{figure2}.
With a convenient choice of the parameters, it is possible to predict an
elastically induced first-order transition from the paraelectric to the
antiferroelectric phase for $x\approx 1$, turning into second order for
smaller values of concentration, and which is in qualitative agreement with
the experimental findings. The glassy transition can still be understood
within the framework of the rigid SK models, with the well-known instability
of the replica-symmetric solution in the low-temperature region. It is
important to remark that the inclusion of Gaussian random fields is not
capable of explaining a first-order transition, in contrast to the
predictions of the proposed compressible model \cite{soares94,vieira00}. In
fact, first-order transitions are associated to either a double-delta
distribution of random fields \cite{nogueira98,vieira00} or to the
consideration of certain discrete quadrupolar-glass and Potts-glass models 
\cite{binder92}.

\section*{Acknowledgement}

We acknowledge useful remarks of Francisco A. da Costa. We also acknowledge the financial support from Conselho Nacional
de Desenvolvimento Cient\'ifico e Tecnol\'ogico (CNPq).

\end{document}